\documentclass[12pt]{article}
 \usepackage{graphicx}
 \usepackage{amssymb,amsmath}
\begin{document}

\title{Transitions between hyperfine structure states of antiprotonic $^4
\mathrm{He}$ at collisions with medium atoms: interaction \emph{ab initio}}
\author{A.V. Bibikov, G.Ya. Korenman, and S.N. Yudin\\
\emph{\small Skobeltsyn Institute of Nuclear Physics,}\\ \emph{\small 
Lomonosov Moscow State University,}\\ \emph{\small Moscow, Russia, 119991}
\small E-mail: korenman@nucl-th.sinp.msu.ru\\
}
\date{}
\maketitle

\begin{abstract}
Collisions of metastable antiprotonic helium atoms with atoms of the medium 
induce, among other processes, transitions between hyperfine structure (HFS) 
states, as well as shifts and broadening of microwave M1 spectral lines.  In 
order to obtain matrix potential of interaction between $(\mathrm{\bar{p} He}
^+)$ and $\mathrm{He}$, we have calculated the  potential energy surface (PES)
 in the framework of unrestricted Hartree-Fock method taking into account 
electron correlations in the second-order perturbation theory (MP2). With 
this potential, the system of close-coupling equations for HFS channels is 
solved numerically. Cross sections and transition rates, shifts and 
broadening of M1 spectral lines are calculated. They are used to obtain a 
numerical solution of the master equation that determines the time evolution 
of the HFS-states density matrix. The results are compared with the 
experimental data and with the results of model calculation.
\end{abstract}

\section{Introduction} \label{sec1}
Since the early 2000s, the ASACUSA collaboration carried out a series of 
experiments on a low-energy antiproton beam at the AD facility in CERN to 
study M1 transitions between sublevels of the hyperfine structure (HFS) 
of long-lived states of antiprotonic atoms $^4\mathrm{He}$ \cite{ref1,ref2} 
and $^3\mathrm{He}$ \cite{ref3} under the influence of microwave (MW) 
irradiation. The experiments were carried out at a low temperature of target 
($T\simeq$6 K) by the triple resonance method. At first, the relative 
population of the lower HFS states is depleted by a laser pulse, then 
the populations of upper and lower groups of HFS states are redistributed by a 
resonance MW radiation, and finally the second laser pulse induces antiproton 
transition from lower HFS sublevels of the metastable state to short-lived 
states that quickly leads to registered annihilation. The main aim of the 
experiments was to measure the HFS splitting and to determine related  
fundamental antiproton characteristics. Along with this, the experiments give 
data on effect of medium on relaxation of HFS populations, collisional shifts 
and broadening of the spectral lines of M1 transitions induced by MW radiation.

In the papers \cite{ref4,ref5}, these effects were considered in the framework of 
the model  interaction between $(\mathrm{\bar{p} He}^+)$ and $\mathrm{He}$ 
that includes scalar and tensor terms of the potential having a 
correct behaviour ($\sim 1/R^6$) at large distances and repulsion at small 
distances ($R\lesssim 1\, a.u.$). With a suitable choice of parameters, this 
model allows to describe the experimental data on the relaxation and collisional 
effects. However, this approach seems insufficient or incomplete due to an 
uncertainty in the choice of possible specific shape of the potential and its 
parameters for different states of $\bar{p}$-atom. 

In this work, interaction between thermalized antiprotonic 
$(\mathrm{\bar{p}}^4\mathrm{He}^{+})$ atom  and ordinary $^4\mathrm{He}$ atom
is described by an \emph{ab initio} potential energy surface (PES) calculated  
in the framework of unrestricted Hartree-Fock method with account for electron 
correlations in the second-order perturbation theory (MP2). In the system 
under consideration, one of the centers (antiproton) has a negative charge. 
Therefore, an application of conventional quantum chemistry methods of PES 
calculations to this system requires some modification and additional tests of 
the calculation accuracy. An analysis of these issues will be presented in a 
more detailed publication. With the obtained potentials, we solve the close
coupling equations in the space of HFS states, find S-matrix and calculate the 
elementary collision characteristics (cross sections and rates of transitions 
between HFS sublevels, shifts and broadening of M1 spectral lines). Then, 
using these values, we solve the quantum kinetic equation (master equation), 
which determines the time evolution of the spin density matrix of HFS states in 
the presence of MW radiation. The results are compared with experimental data 
\cite{ref1,ref2} and model calculations. 

\section{Potential energy surface and $\mathbf{(\bar{p}He^+) - He}$ interaction
 potentials} \label{sec2}

The system consists of three heavy particles and three electrons. Denote by 
$m_\mathrm{\bar{p}}$, $M_a$, $M_b$ and $\mathbf{r}_\mathrm{\bar{p}}$, 
$\mathbf{R}_a$, $\mathbf{R}_b$ masses and coordinates of antiproton, 
nucleus $a$ (in $ \bar{p}$-atom) and helium atom, respectively. The coordinates 
of He atom and of nucleus $b$ coincides with the accuracy of order $m_e/M_b$.
Let us introduce Jacobi coordinates of heavy particles: $\mathbf{r}=
\mathbf{r}_\mathrm{\bar{p}}-\mathbf{R}_a$ and 
$\mathbf{R}=\mathbf{R }_b - (\lambda \mathbf{r}_\mathrm{\bar{p}}+\nu \mathbf{R}_a)$,
where $\lambda =  M_a/(M_a + m_\mathrm{\bar{p}})$, 
$\nu = m_\mathrm{\bar{p}}/(M_a + m_\mathrm{\bar{p}})$. 
Heavy particles move much slower than electrons, so we can use an adiabatic 
approximation and present the interaction energy between antiprotonic and 
ordinary atoms as
 \begin{equation} \label{eq1}
 V(r, R, \cos\theta) = 4/|\mathbf{R} + \nu \mathbf{r}| -  
2/|\mathbf{R} - \lambda \mathbf{r}| + E_e(r, R, \cos\theta) - E(\mathrm{He
}) - E(\mathrm{\bar{p}He}^+), 
\end{equation} 
where two first terms are Coulomb interactions of nucleus $b$ 
with nucleus $a$ and antiproton, $E(\mathrm{He})$ and $E(\mathrm{\bar{p}He}^+)$ 
are internal energies of isolated subsystems, 
$\cos\theta=(\mathbf{R\cdot r})/Rr$. The energy of three electrons 
$E_e(r, R, \cos\theta)$, as well as internal energies of antiprotonic
and ordinary atoms were calculated in unrestricted Hartree-Fock 
approximation taking into account electron-electron correlations in the 
second-order perturbation theory. An extended set of molecular basis functions 
aug-cc-pV5Z \cite{ref6} was used, taking into account correlations and valence 
polarization, with parameters from \cite{ref7}. Electronic orbitals were 
centred on $a$ and $b$ nuclei.
Numerical calculations were performed using an original program based on the RI 
("resolution of identity") method for computing of the 
integrals of electron-electron interactions, which significantly reduces the 
computational cost in Hartree-Fock approximation when considering large 
systems or systems with a large number of configurations \cite{ref8,ref9}.

To highlight the angular dependence of $V(r,R,\cos\theta)$ we expand it in a 
series of Legendre polynomials 
\begin{equation} \label{eq2}
V(r,R,\cos\theta) = \sum^\infty_{k=0} V^k(r,R) P_k (\cos\theta), 
\end{equation}
where
\begin{equation} \label{eq3}
V^k(r,R) =(k+1/2)\int^{1}_{-1}V(r,R, t)P_k (t)dt.
\end{equation}
The performed calculations show that Eq. \eqref{eq1} at $R\gtrsim r$ depends 
weakly on $\cos\theta$, therefore for considering of thermal collisions, we can
restrict the series \eqref{eq2} by the lowest multipoles. 

Antiproton quantum numbers $n,L$ do not change in the transitions between HFS 
states, therefore it is convenient to introduce averaged values
\begin{equation} \label{eq4}
V^k_{nL}(R) = \int^\infty_0V^k(r,R) u^2_{nL}(r)r^2dr,
\end{equation}
where $u_{nL}(r)$ is a radial wave functions of antiproton, which is calculated 
within the same approximations as the energy $E(\mathrm{\bar{p}He}^+)$ in Eq.
\eqref{eq1}. 

Splitting of the levels of $(\mathrm{\bar{p}^4He}^+)_{nL}$ atom into 4 HFS
sublevels arises due to interaction of magnetic momenta associated with 
orbital angular momentum $\mathbf{L}$ and spins of electron ($\mathbf{s}_e$) 
and of antiproton ($\mathbf{s}_{\bar{p}}$). It follows from the calculations 
\cite{ref10,ref11} that the HFS states can be characterized approximately by 
quantum numbers $F=L\pm s_e$,  $J=F\pm s_{\bar{p}}$. Corresponding spin-angle 
functions $|Ls_e(F)s_{\bar{p}}JM\rangle$ are obtained by the vector coupling 
of the momenta $\mathbf{L} +\mathbf{s}_e=\mathbf{F},\quad 
\mathbf{F} + \mathbf{s_{\bar{p}}}=\mathbf{J}$.
For brevity sake, we enumerate these states at fixed $n,\,L$ in accordance 
with the energy position  ($\epsilon_1<\epsilon_2<\epsilon_3<\epsilon_4$),
\begin{equation} \label{eq5}
\begin{split}
|1\rangle &= |F=L+1/2, J=F-1/2=L \rangle, \\
|2\rangle &= |F=L+1/2, J=F+1/2=L+1 \rangle, \\
|3\rangle &= |F=L-1/2, J=F-1/2=L-1 \rangle, \\
|4\rangle &= |F=L-1/2, J=F+1/2=L \rangle. \\
\end{split}
\end{equation} 
Matrix of potentials for transitions between HFS states in collisions of 
$\bar{p}$-atom with $^4\mathrm{He}$ atom with account for lowest 
multipoles up to $k=2$ in Eq. \eqref{eq2} can be written as 
 \begin{equation} \label{eq6}
V_{cc'}(R) = V^0_{nL}(R) \delta_{cc'}+V^2_{nL}(R) \cdot 
\langle L s_e(F) s_{\bar{p}}(J)l:j|P_2(\cos\theta)| L s_e(F') s_{\bar{p
}}(J')l':j\rangle
\end{equation}
where multi-index $c$  includes quantum numbers of the state ($n,L,F,J$), 
as well as orbital angular momentum $l$ of the relative motion of subsystems 
and total angular momentum $j$ of the whole system. The spin-angular matrix 
element in \eqref{eq6} is expressed by standard methods in terms of $3j$ and 
$6j$-symbols. Matrix potential \eqref{eq6} has the same structure as the
model \cite {ref4, ref5}, but functions $V^0_{nL}(R)$ and $V^2_{nL}(R)$ can 
differ from model ones.
Radial dependence of monopole terms of model and \emph{ab initio} potentials for
interaction between $(\bar{p}\mathrm{He}^+)_{37,35}$ and ordinary $^4\mathrm{He}$
atom is shown in Fig. 1. At large distances, these potentials coincide, but 
the repulsion and minimum regions of the model potential are shifted toward 
smaller distances, and the well depth is greater than that of the 
\emph{ab initio} potential. 

\begin{figure}[thb]
\begin{center}
	\includegraphics[width=0.75\textwidth]{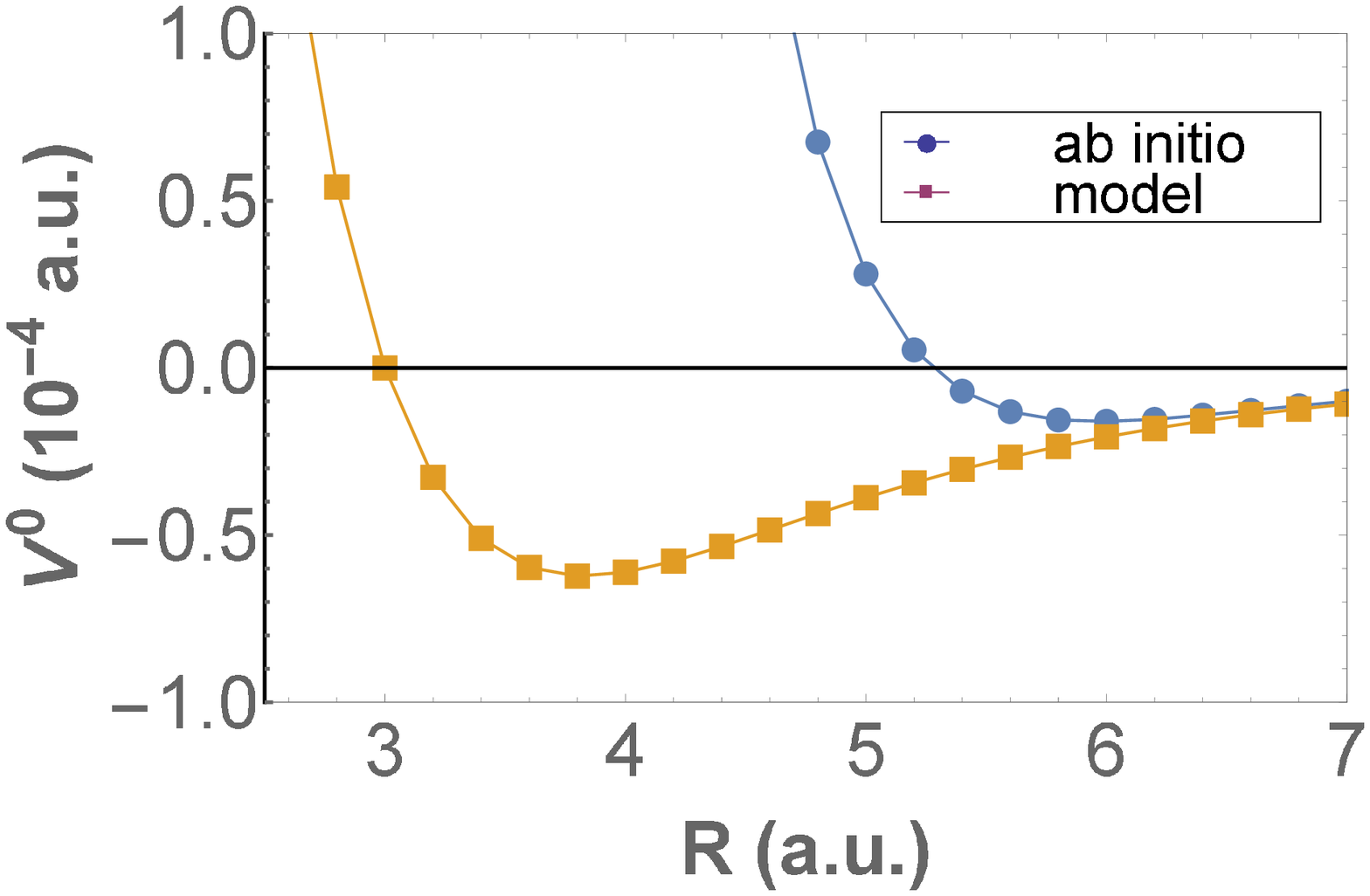}
	\caption{Radial dependence of monopole terms of model and \emph{ab initio} 
potentials for interaction between $(\bar{p}\mathrm{He}^+)_{37,35}$ and 
ordinary $^4\mathrm{He}$ atom.}
	\end{center}
	\label{fig1}
	\end{figure}

\section{Elementary characteristics of collisions} \label{sec3}

To obtain cross sections of transitions between HFS states, shifts and 
broadening of M1 lines induced by MW pulse, we solve numerically a quantum 
problem of coupled channels with different $F,J$ at fixed $n,L$. 
After separating the spin and angular variables, the system of equations for 
radial wave functions $Z_c(R)$ of relative motion of the subsystems takes the 
following form
\begin{equation}\label{eq7}
Z_c''(R)+[k_c^2+l_c(l_c+1)/R^2]Z_c(R) = 2M\sum_{c'}V_{cc'}(R)Z_{c'}(R),
\end{equation}
where $k_c=\sqrt{2M(E +\epsilon_i-\epsilon_c)}$, $M$ is a reduced mass of 
colliding subsystems, $E$ is a kinetic energy in input channel, $\epsilon_i$ 
and $\epsilon_c$ are energies of HFS sublevels in the input channel 
$i$ and in a channel $c$. Matrix of the potentials $V_{cc'}(R)$ is 
determined by Eq. \eqref{eq6}.

Cross sections and transition rates were calculated using standard formulas
\begin{align}
\sigma(FJ\rightarrow F'J') & =\frac{\pi}{k_i^2} \sum_{jll'}
\frac{2j+1}{2J+1} \cdot\left|\delta_{FF'} \delta_{JJ'} \delta_{ll'} -
\langle F'J'l'|S^j|FJl\rangle \right|^2,
 \label{eq8} \\
\label{eq9}
\lambda(FJ\rightarrow F'J') & =  N\big\langle v \sigma(FJ\rightarrow F'J') 
\big\rangle,
\end{align}
where $N$ is atomic density of the medium; the outer angular brackets in 
\eqref{eq9}  mean averaging over thermal motion of the colliding atoms.
Shift and broadening of M1 line for $F_1 J_1\rightarrow F_2 J_2$ transition 
can be considered according to \cite{ref12}, Eq. (59.98). In our notations, this 
equation has the following form: 
\begin{multline}\label{eq10}
\gamma + \mathrm{i}\Delta = N\pi \sum_{ll'j_1j_2}
(2j_1+1)(2j_2+1)(-1)^{l+l'} \cdot \left\{\begin{matrix} j_1 & j_2 & 1 \\ J_2 &
 J_1 & l \end{matrix}\right\}
\left\{\begin{matrix} j_1 & j_2 & 1 \\ J_2 & J_1 & l'\end{matrix}\right\} 
\\ \cdot  \big\langle
vk^{-2}\bigl[ \delta_{ll'} -  \langle nLF_1J_1 l'|S^{j_1}_I|nLF_1J_1 l \rangle
 \cdot  \langle nLF_2J_2 l'|S^{j_2}_{II}|nLF_2J_2 l\rangle ^* \bigr] 
\big\rangle  ,
\end{multline}
where $S_I$- and $S_{II}$-matrixes refer to collisions before and after M1 
transition $F_1 J_1 \rightarrow F_2 J_2$.
Table \ref{tab1} shows transition rate constants $\langle\sigma v\rangle$, 
per-atom shifts $\Delta/N$ and broadenings $\gamma/N$ for transition 
$2 \rightarrow 4$ ($n,L=37,35$), calculated with model potential \cite{ref4} 
and \emph{ab initio} potential 
\eqref{eq5}, at the medium temperature $T=6\,K$. It is seen that 
the values obtained with \emph{ab initio} potential are several times smaller 
than those obtained with the model potential. The main reason for this 
difference is related to the difference in the behaviour of potentials noted 
above in the discussion of Fig. \ref{fig1}.
The rates of collisional transitions between HFS states are not directly 
measured in the existing experiments. The calculated shifts of M1 lines are very 
small that does not contradict to data in the Refs. \cite{ref1, ref2}, according to 
which the shifts are much less than the measurement accuracy of the line
frequency. 
The value of the collisional broadening of the line constitutes only a small 
part of the observed width. The main contribution to the width is provided by  
'Fourier broadening', determined by a finite time of the microwave irradiation 
\cite{ref1, ref2}. However, the mentioned experiments to measure HFS splitting
give also dependencies of the relative magnitude of the annihilation signal on
the MW frequency and on the delay time of second laser pulse. 
For a theoretical description of these quantities, let us consider time 
evolution of the spin density matrix of HFS states and its dependence on the MW
frequency.

\begin{table}
\begin{center}
\caption{Transition rate constants $\langle\sigma v\rangle$ and per-atom shifts 
$\Delta/N$ and broadenings $\gamma/N$ for transition 
$2\rightarrow 4$ ($n,L=37,35$) with model and \emph{ab initio} potentials at 
$T=6\, K$. All quantities are given in atomic units.} \label{tab1}
\vspace{2mm}
\begin{tabular}{lccc}\hline
	& $\langle\sigma v\rangle$ $\times$ $10^7$ & $(\Delta/N)\times 10^9$ & $(\gamma/N)\times 10^7$\\ \hline
	model & 4.46 & 40.53 & 12.03\\
	ab initio & 1.11 & 6.64 & 2.03\\ \hline
\end{tabular}
\end{center}
\end{table}

\section{Time evolution of spin density matrix of HFS states} \label{sec4}

The basic quantum kinetic equation (Master Equation) \cite{ref13} for our 
task can be represented as follows
\begin{equation} \label{eq11}
\frac{d\rho_{ij}(t)}{dt} =
-(\mathrm{i}\omega_{ij}+\lambda_r)\rho_{ij}(t) - \mathrm{i}
\left[V(t),\rho(t) \right]_{ij} + \sum_{km} R_{ijkm} \rho_{km}(t) 
+\delta_{ij}\beta_i,
\end{equation}
where indices $i,j,k,m$ are the numbers of HFS states of the system, 
$\omega_{ij}=\epsilon_i - \epsilon_j$,  
$\lambda_r$ is the spontaneous (radiative) decay rate of $(n,L)$-state of the 
antiproton atom, $V(t)$ is an external microwave field, $\beta_i$ is a 
refilling rate of HFS state due to transitions from higher states to $(n,L)$ 
level. The relaxation term on the right-hand side \eqref{eq11}, containing matrix
$R_{ijkm}$, takes into account effects of collisions between  $\bar{p}$-atom and
atoms of the medium on an evolution of the density matrix of HFS states. In the 
secular approximation \cite{ref13}, nonzero elements of the relaxation matrix 
are only those that satisfy the condition $\omega_{ij}=\omega_{km}$ that leads 
to time independence of the relaxation matrix. Thus, in short form:
\begin{equation} \label{eq12}
R_{ijkm}= - \lambda_i\delta_{ij}\delta_{km}\delta_{ik} +
\lambda(k\rightarrow i)\delta_{ij}\delta_{km}(1-\delta_{ik}) - (\gamma_{ij} + 
\mathrm{i}\Delta_{ij})\delta_{ik}\delta_{jm}(1-\delta_{ij}),
\end{equation}
where $\lambda_i = \sum_{f\neq i} \lambda(i\rightarrow f)$. Substituting 
\eqref{eq12} into \eqref{eq11}, we make sure that 
$(\Delta_{ij}- \mathrm{i}\gamma_{ij})$ can be treated as a complex shift of the 
frequency  $\omega_{ij}$, and $\gamma_{ij}$ provides line 
broadening $i\rightarrow j$ due to collisions. 

Interaction of the system with an external MW field has the following form:
\begin{equation} \label{eq13}
V_{ij} (t)=-(\mu_z )_{ij} B_0 \cos\omega t,
\end{equation}
where $\mu_z$ is an magnetic moment operator of the $\bar{p}$-atom,  and $B_0$ 
is a magnetic field intensity. Assuming that the frequency detuning is small 
($|\omega-\omega_{ij}|\ll \omega$) 
we use the "rotating field" approximation \cite{ref13} in Eq. \eqref{eq11}, that
allows to neglect by rapidly oscillating terms in interaction representation,
and, in particular, to omit the diagonal components of interaction with the 
alternating field.

Until the first laser pulse, the relative populations of HFS sublevels are 
proportional to their statistical weights $(2J+1)/4(2L+1)\simeq 1/4$ (for $L
\gg 1) $. After that, at the initial time, two lower sublevels are depleted 
by $\varepsilon$, and the off-diagonal elements of the density matrix remain 
zero. At this moment, the action of MW radiation begins. Therefore, we can 
take initial conditions for Eq. \eqref{eq11} as
\begin{equation} \label{eq14}
\rho_{ij}(t=0) =(1/4) \delta_{ij} \big[1-\varepsilon(\delta_{i1} + \delta_{i2}
) \big]. 
\end{equation} 
Value $\beta_i$ is also proportional to the statistical weight of HFS state, 
$\beta_i\simeq \beta_{tot}/4$. We assume that the full refilling rate 
 of ($n,L$)-state  is equal to the rate of its radiative decay, 
$\beta_{tot}=\lambda_r$ that compatible qualitatively with indirect 
experimental results. In the calculations for state $(n,L)$= (37,35), 
value $\lambda_r = 7.15 \times 10^5$ $s^{-1}$ \cite{ref14} was used.

\begin{figure}[thb]
\begin{center}
	\includegraphics[width=0.75\textwidth]{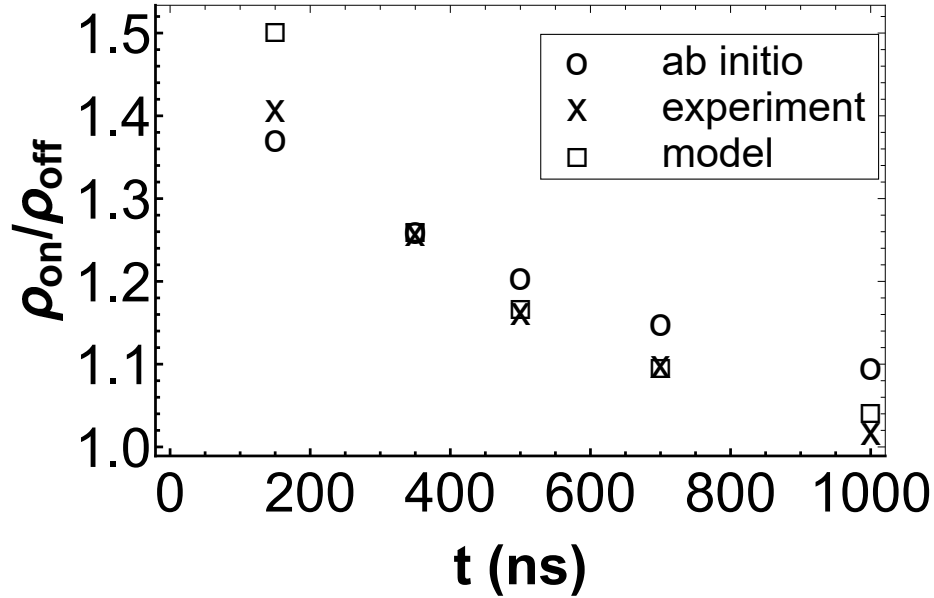}
	\caption{Dependence of relative annihilation signal on delay time 
of second laser pulse from the calculations with model and \emph{ab initio} 
potentials, and from experimental data.}
\end{center}
	\label{fig2}
\end{figure} 
  
Fig. 2 shows dependence of the relative values of annihilation signals on 
the time of the second laser pulse from the calculations with model and \emph{ab initio} interaction potentials as well as from experimental data. The 
degree of depopulation $\varepsilon$ by the primary laser pulse was used as a 
free parameter. It was chosen from the condition that the calculated 
relative value of the annihilation signal is equal to experimental one at a 
delay time of the second laser pulse of $t = 350\, ns$, corresponding to the 
best experimental statistics. We adopt $\varepsilon = 0.65$ and $\varepsilon= 
0.5$ in the calculations with model and \emph{ab initio} potentials, respectively.
All calculated values agree fairly well with the experimental data, although 
the results of model calculations for the delay time dependence are in general
somewhat closer to the measured ones due to free fitting model parameters.

\section{Conclusion} \label{sec5}

Application of potential energy surface to interaction between  
$(\mathrm{\bar{p} He}^+)$ and $\mathrm{He}$ atoms allows to consider 
effects of collisions on the transitions between HFS states of $\bar{p}$-atom in
a low-temperature medium without introducing a model interaction that involves a
controversial choice of some detailed form and parameters of the potential. 
Moreover, the approach using PES allows to consider without model also transitions of
$\bar{p}$-atom with change of quantum numbers $n,L$, especially most interesting
collisional Stark transitions $L\rightarrow L'$.

The results shown above refer to HFS states at $n,L=37,35$ that were studied 
experimentally in most details. Results for different $n,L$ and for two 
isotopic targets ($^4\mathrm{He}$ and $^3\mathrm{He}$), as well as a detailed 
comparison with experimental data will be published separately.

One of the authors (G.K.) is thankful to T. Yamazaki, R. Hayano, and E. 
Widmann for drawing our attention to the issues under consideration and for 
helpful discussions.

\end{document}